%% file: iwaenc.tex
\documentclass{article}
\usepackage{spconfa4,amsmath,graphicx}
\usepackage[usenames]{color}
\usepackage{mathtools}
\usepackage{xargs}
\usepackage{amsfonts}
\usepackage{amsmath}
\usepackage{algorithmic}
\usepackage{amssymb}
\usepackage{pifont}
\usepackage{stylesheet}
\usepackage{todonotes}
\usepackage{pgfplots}
\usepackage{soul}
\usepackage{xcolor}
\usepackage{accents}

\DeclareUnicodeCharacter{2212}{−}
\usepgfplotslibrary{groupplots,dateplot}
\usetikzlibrary{patterns,shapes.arrows,external}
\pgfplotsset{compat=newest}

\title{ONE-SHOT DISTRIBUTED NODE-SPECIFIC SIGNAL ESTIMATION WITH NON-OVERLAPPING LATENT SUBSPACES IN ACOUSTIC SENSOR NETWORKS}

\name{Paul Didier, Pourya Behmandpoor, Toon van Waterschoot, Marc Moonen\thanks{This research was carried out at the ESAT Laboratory of KU Leuven, in the frame of Research Council KU Leuven C14-21-0075 ``A holistic approach to the design of integrated and distributed digital signal processing algorithms for audio and speech communication devices'', supported by the European Union's Horizon 2020 research and innovation programme under the Marie Skłodowska-Curie grant agreement No. 956369: `Service-Oriented Ubiquitous Network-Driven Sound — SOUNDS', and in the frame of Research Project FWO nr. G0C0623N 'User-centric distributed signal processing algorithms for next generation cell-free massive MIMO based wireless communication networks'.
The scientific responsibility is assumed by its authors. This paper reflects only the authors' views and the Union is not liable for any use that may be made of the contained information.}}
\address{
    KU Leuven, Department of Electrical Engineering (ESAT),\\
    STADIUS Center for Dynamical Systems, Signal Processing and Data Analytics, Belgium
}
\begin{document}
\maketitle
\begin{abstract}
A one-shot algorithm called iterationless DANSE (iDANSE) is introduced to perform distributed adaptive node-specific signal estimation (DANSE) in a fully connected wireless acoustic sensor network (WASN) deployed in an environment with non-overlapping latent signal subspaces. The iDANSE algorithm matches the performance of a centralized algorithm in a single processing cycle while devices exchange fused versions of their multichannel local microphone signals.
Key advantages of iDANSE over currently available solutions are its iterationless nature, which favors deployment in real-time applications, and the fact that devices can exchange fewer fused signals than the number of latent sources in the environment. The proposed method is validated in numerical simulations including a speech enhancement scenario.
\end{abstract}
\begin{keywords}
Speech enhancement, distributed signal estimation, sensor networks, latent subspaces
\end{keywords}
\section{Introduction}
\label{sec:intro}

The recent increase in acoustic sensor availability enables the creation of wireless acoustic sensor networks (WASNs), which can surpass the individual device performance and be more flexible than centralized systems~\cite{plata2017heterogeneous}.
Their use relies on distributed algorithms with low latency and low bandwidth usage.
Most existing solutions focus on the estimation of parameter vectors~\cite{cobos_survey_2017,zhao2020model} or a single globally defined quantity~\cite{schizas_distributed_2007}, even though many applications require estimation of node-specific target signals~\cite{bertrand2011applications}. The exploration of this problem has led to the distributed adaptive node-specific signal estimation (DANSE) algorithm~\cite{bertrand2010distributed} and variants thereof~\cite{bertrand2011distributed, szurley2016topology,plata2015distributed}. If the target sound sources are all observable by all nodes, their latent subspaces fully overlap and DANSE reaches the performance of an equivalent centralized algorithm. Remarkably, DANSE reaches optimality while nodes only exchange dimensionally reduced (so-called fused) versions of their local sensor signals.

In general, however, the target signals of different nodes may span different latent subspaces which may not mutually overlap.
For example, signals recorded in an airport or a train station may exhibit non-overlapping latent subspaces if some sound sources produce signals that are of interest to node $k$ but that cannot be captured by node $q$. This may be the case although $k$ and $q$ can exchange data and may simultaneously aim at estimating a signal stemming from a shared fully overlapping latent subspace, e.g., a public address (PA) signal.

This scenario has been addressed in~\cite{plata2015distributed}, where it was concluded that each node should broadcast as many fused signals as the number of latent sources observed by the entire WASN for DANSE to remain optimal.
Even with such increased communication bandwidth, DANSE still requires several iterations to converge, a clear limiting factor for applications requiring real-time processing~\cite{szurley2016topology}. To address both of these shortcomings, we present the iterationless DANSE algorithm (iDANSE), which matches the centralized signal estimation performance in a single processing cycle (i.e., in one shot) even if non-overlapping latent subspaces are present. Unlike in~\cite{plata2015distributed}, nodes within iDANSE do not need to exchange more fused signals than the dimension of the fully overlapping latent subspace shared by all nodes.

\section{Problem Statement}
\label{sec:prob_statement}

\subsection{Signal model}\label{subsec:sig_model}
\noindent
Consider a WASN with $K$ nodes and $M_k$ sensors at node $k$, where $k\in \K \coloneqq \{1,\dots,K\}$ and $M \coloneqq \sum_{k\in\K} M_k$.
The WASN is deployed in a scene composed of $S$ desired sources and $N$ noise sources, all mutually uncorrelated.
A source is said to be ``global'' if its signal is captured by all nodes (e.g., a PA system) or ``local'' if its signal is only captured by one node (e.g., a nearby conversation partner). The proposed algorithm can be extended to include cases where a latent signal is captured by more than one but not all nodes, although this is not exposed here.
Among the $S$ desired sources, $\cS$ are global, grouped in the set $\cSset$. The other $\uS \coloneqq S - \cS$ sources are local, grouped in $\uSset$. This notation (``circle $\circ$'' or ``dot $\cdot$'') is adopted for clarity as the dot represents a single node and the circle the whole WASN.
For any $k\in\K$, there exist $\uSk$ local desired sources, grouped in $\uSsetk$, that contribute to its local sensor signals.
Analogous notation is adopted for the noise, replacing $S$ and $\Sset$ by $N$ and $\Nset$, respectively.
This results in the signal model for the $M_k$ local sensor signals $\cy_k$ of node $k$:
\begin{equation}\label{eq:basisSigModel}
    \cy_k \coloneqq \cs_k + \cn_k \coloneqq \ccs_{k} + \cus_{k} + \ccn_{k} + \cun_{k}~\in\C^{M_k},
\end{equation}
where $\cs_k \coloneqq \ccs_k + \cus_k$ is the desired signal component with $\ccs_k$ and $\cus_k$ the global and local contributions, respectively, and analogous notation is used for the noise $\cn_k \coloneqq \ccn_k + \cun_k$. Note that sensor noise is included in $\cun_k$.
Signals are assumed complex-valued to allow frequency-domain representations.

\begin{figure}[t!]
    \centering
    \includegraphics[width=.8\columnwidth,trim={0 0 0 0},clip=false]{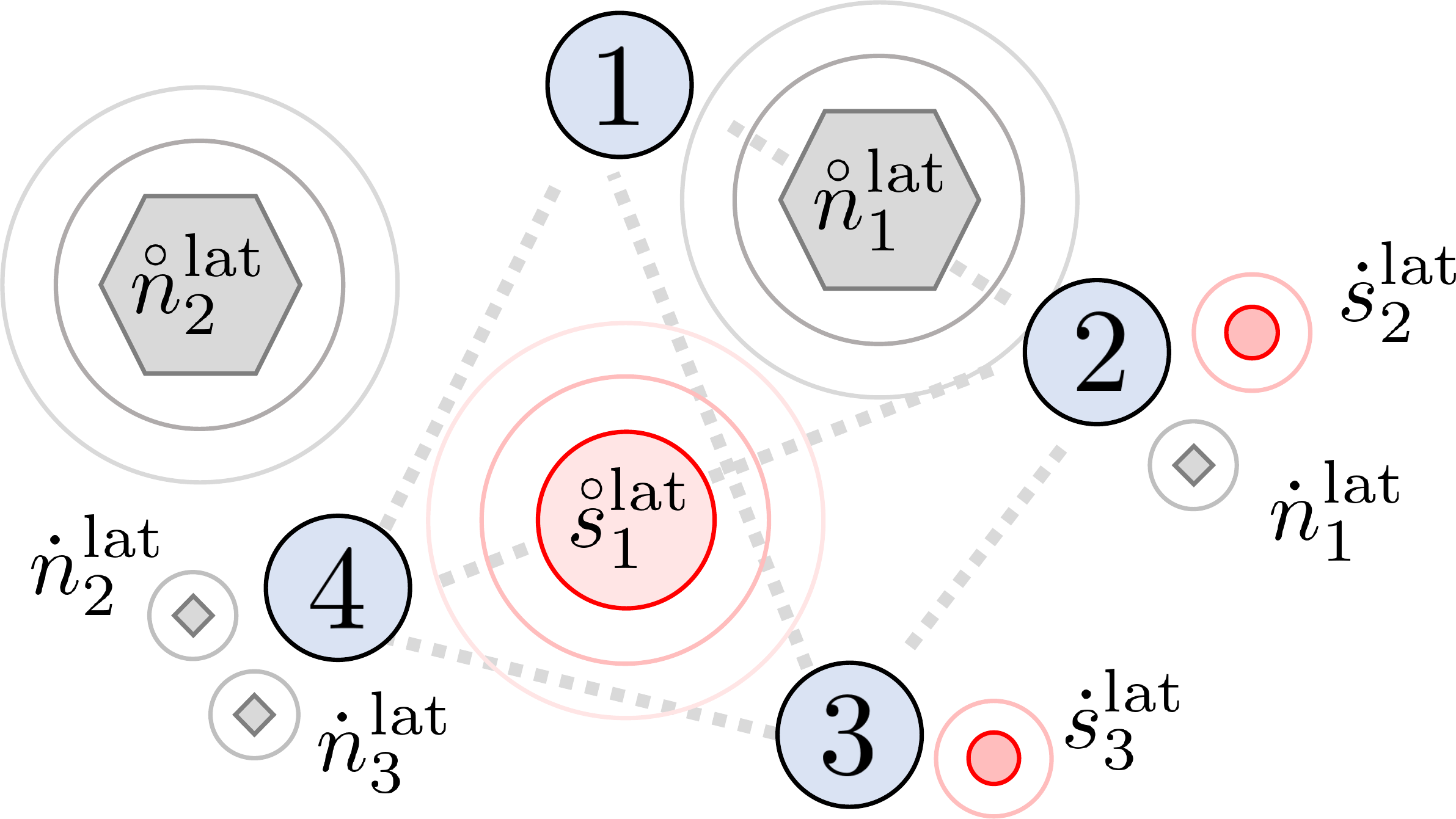}
    \caption{Scenario with 4 nodes, 1 global and 2 local desired sources (red), 2 global and 3 local noise sources (grey).}
    \label{fig:scenario}
\end{figure}

All local sensor signals can be stacked into a centralized $M$-dimensional vector $\cy = \cs + \cn = \ccs + \cus + \ccn + \cun$, where $\cy \coloneqq [\ly[\T][1] \dots \ly[\T][K]]^\T$ with $\cdot^\T$ the transpose operator, and the other vectors are analogously defined from~\eqref{eq:basisSigModel}.
The relevant terms in this centralized model can be further expanded as:

\vspace*{-.75em}

\begin{equation}\label{eq:centrSigModel}
    \cy =
    \cA\ccs[\mathrm{lat}]
    + \uA\cus[\mathrm{lat}]
    + \ccn
    + \cun
    \in\C^M,
\end{equation}
where
$\cA \in \C^{M \times \cS}$ and $\uA \in \C^{M \times \uS}$ are steering matrices between the $M$ sensors and the sources in $\cSset$ producing the latent signals $\ccs[\mathrm{lat}]$ and those in $\uSset$ producing $\cus[\mathrm{lat}]$, respectively. An example is depicted in~\figref{fig:scenario}, where the same notation is adopted for noise latent signals $\ccn[\mathrm{lat}]$ (global) and $\cun[\mathrm{lat}]$ (local).
The matrix $\uA$ has a sparse block structure since any node $k$ only captures the signals of the local sources in $\uSsetk$.
Let the signals in $\ccs[\mathrm{lat}]$ be ordered from the latent signals generated by the sources in $\uSsetk[1]$ down to $\uSsetk[K]$. Then $\uA$ can be written as:

\begin{equation}\label{eq:LocalSteeringMat}
    \uA \coloneqq \begin{bmatrix}
        \uA_1\\
        \vdots\\
        \uA_K
    \end{bmatrix}
    \coloneqq \begin{bmatrix}
        \uA_1' & \dots & \mathbf{0}\\
        \vdots & \ddots & \vdots\\
        \mathbf{0} & \dots & \uA_K'
    \end{bmatrix},
\end{equation}
where $\uA_k'\in\C^{M_k\times \uSk}$ is the non-zero part of $\uA_k\in\C^{M_k\times \uS}$.

\subsection{Centralized solution}
\label{subsec:centr}
\noindent
Consider the centralized case where each node $k\in\K$ aims at estimating a $J$-channel target signal $\ld \coloneqq \Ekk^\T\cs_k$, where $\Ekk$ is an $M_k\times J$ selection matrix. The value of $J$ is assumed equal for all nodes without loss of generality, although it may in princple differ between nodes. Node $k$ solves the linear minimum mean square error (LMMSE) problem:

\vspace*{-1em}

\begin{equation}\label{eq:mmseCentr}
    \mathbf{W}_k \coloneqq \underset{\mathbf{W}\in \C^{M\times J}}{\mathrm{arg\,min}}\:
    \E[{\|\ld - \mathbf{W}^\Her\mathbf{y}\|^2_2}],\:\forall\:k\in\K,
\end{equation}

\noindent
where $\E[{\cdot}]$ denotes the expectation operator, $\|\cdot\|_2$ the Euclidean norm, and $\cdot^\Her$ the Hermitian transpose. The solution to this centralized signal estimation problem is the multichannel Wiener filter (MWF):

\begin{equation}\label{eq:mwfCentr}
    \mathbf{W}_k=\left(\mathbf{R}_{\mathbf{yy}}\right)^{-1}\Rydk,\:\forall\:k\in\K,
\end{equation}

\noindent
with the spatial covariance matrices (SCMs) $\mathbf{R}_{\mathbf{yy}}\coloneqq\E[{\cy\cy[\Her]}]$ and $\Rydk\coloneqq\E[{\cy\ld[\Her]}]$. The target signal is then estimated as $\hd \coloneqq \mathbf{W}_k^\Her\cy$. Note that, since the desired and noise sources are uncorrelated, $\Rydk \coloneqq \mathbf{R}_{\mathbf{ss}}\mathbf{E}_{k}$ where $\mathbf{R}_{\mathbf{ss}} \coloneqq \E[{\cs\cs[\Her]}]$ and $\mathbf{E}_{k}$ is an $M\times J$ matrix extracting $\ld$ from $\mathbf{s}$. 

Note that $\mathbf{R}_{\mathbf{yy}}$ can be directly estimated from the microphone signals via averaging techniques as described in, e.g.,~\cite[Sec.V-A]{bertrand2010distributed}. In the case of speech as desired signal, a voice activity detector (VAD) can be obtained using various methods~\cite{zhao2020model, dov2017multimodal, kim2018voice} and used to compute $\mathbf{R}_{\mathbf{ss}}$ as $\mathbf{R}_{\mathbf{yy}} - \mathbf{R}_{\mathbf{nn}}$.

\subsection{Review of the DANSE algorithm}
\noindent
The DANSE algorithm~\cite{bertrand2010distributed} is an iterative distributed algorithm that can be used to solve the estimation problem \eqref{eq:mmseCentr} in a fully connected WASN. At DANSE iteration $i$, node $k$ defines a so-called fusion matrix $\Pk^i\in\C^{M_k\times J}$ to compute the fused signal $\oz[i] \coloneqq \Pk^{i\Her}\cy_k$. Each node broadcasts its fused signal to all other nodes. Node $k$ then has access to:
\begin{equation}\label{eq:observationVector}
    \oy[i] \coloneqq \begin{bmatrix}
        \ly[\T] & \oz[i\T][-k]
    \end{bmatrix}^\T \in \C^{\tM},
\end{equation}
where $\oz[i][-k] \coloneqq [\oz[i\T][1] \:\dots\:\oz[i\T][k-1]\:\oz[i\T][k+1]\:\dots\:\oz[i\T][K]]^\T$ and $\tM \coloneqq M_k + J(K - 1)$. In practice, iterations are spread over time, with a different frame of signal $\ly$ at each $i$.
Any updating node $k$ solves the following LMMSE problem:

\vspace*{-1.25em}

\begin{equation}\label{eq:mmseDANSE}
    \begin{split}
        \oW[i+1] &\coloneqq \underset{\mathbf{W} \in \C^{\tM \times J}}{\mathrm{arg\,min}}\:
        \E[{\|\ld - \mathbf{W}^\Her\oy[i]\|^2_2}]\\
        &= (\oRyy[i])^{-1} \tilde{\mathbf{R}}_{\mathbf{y}_k\ld}^i,
    \end{split}
\end{equation}

\noindent
where $\oRyy[i] \coloneqq \E[{\oy[i]\oy[i\Her]}]$ and
$\tilde{\mathbf{R}}_{\mathbf{y}_k\ld}^i \coloneqq \E[{\oy[i]\ld[\Her]}]$. These SCMs must be estimated using, e.g., the techniques mentioned in~\secref{subsec:centr}.
The target signal estimate is then obtained as $\hd[i+1] \coloneqq \oW[i+1,\Her]\oy[i]$.
Nodes may use~\eqref{eq:mmseDANSE} to update sequentially~\cite{bertrand2010distributed}, simultaneously, or asynchronously~\cite{bertrand2010distributed2}.
The filters $\oW[i+1]$ can be partitioned as $[\Wkk[i+1,\T]\:|\:\Gkmk^{i+1,\T}]^\T$, where $\Wkk[i+1]\in\C^{M_k\times J}$ is applied to $\cy_k$.
An iteration is completed by setting $\Pk^{i+1} = \Wkk[i+1]$.

\subsection{Discussion on DANSE optimality}\label{sec:DANSEoptConds}

The DANSE algorithm reaches optimality when $\ld$ spans the entire latent subspace defined by the sources that contribute to it.
A too low $J$ leads to an ill-posed problem~\cite[Sec.II-B]{bertrand2010distributed}. For~\eqref{eq:centrSigModel}, supposing $J$ equal for all nodes, this implies:

\begin{equation}\label{eq:full_subspace_span_condition}
    J\geq \cS + \underset{k\in\K}{\mathrm{max}}\:\uSk,
\end{equation}
since the desired latent subspace may contain contributions from both global and local sources.

The condition~\eqref{eq:full_subspace_span_condition} is necessary but not sufficient to ensure convergence. In fact, extensive simulations have shown that DANSE does not reach the centralized solution when local desired sources and global noise sources are present, i.e., when $\uSk > 0$ for at least one node $k$ and $\cN>0$, even if~\eqref{eq:full_subspace_span_condition} holds. This is stated as an observation since a formal convergence proof in those conditions is not available.

\section{The \lowercase{i}DANSE Algorithm}
\label{sec:idanse}

Iterations may be avoided for distributed signal estimation while exchanging fused signals between nodes. \secref{subsec:oneshotexample} provides an intuition for this fact through a mathematical proof for a simple signal model. Extending this reasoning to~\eqref{eq:centrSigModel}, it can be shown that iDANSE (defined in~\secref{subsec:algoDef}) is optimal in one-shot, though the full proof is ommitted.

\subsection{One-shot estimation: an example}\label{subsec:oneshotexample}

Suppose $\cS{\geq}1$, $\cN{=}0$, $\uS{=}0$, and $J{=}\cS$, i.e.,~\eqref{eq:centrSigModel} reduces to $\cy = \cA\ccs[\mathrm{lat}] + \cun$. Since $\ccs[\mathrm{lat}]$ and $\cun$ are uncorrelated, $\Rydk = \cA\cRsslat\cA^\Her\mathbf{E}_k$ where $\cRsslat\coloneqq \E[{\ccs[\mathrm{lat}]\ccs[\mathrm{lat},\Her]}]$.
Furthermore, $\mathbf{R}_\mathbf{yy}$ can be separated as $\cA\cRsslat\cA^\Her + \uRnn$ where $\uRnn\coloneqq \E[{\cun\cun[\Her]}]$ is block-diagonal.
Making use of the matrix inversion lemma, we can re-write~\eqref{eq:mwfCentr} as (details omitted):

\vspace*{-1.5em}

\begin{align}\label{eq:mad}
    \mathbf{W}_k
    &=
    \iuRvv\cA\Tk,\\
    \text{with}\:\:\:
    \Tk &\coloneqq \left(
        \mathbf{I}_J
        -
        \mathbf{X}^{-1}
        \cA^\Her
        \iuRvv
        \cA
    \right)\cRsslat\cA^\Her\mathbf{E}_k,\\
    \mathrm{and}\:\:\:
    \mathbf{X} &\coloneqq (\cRsslat)^{-1} + \cA^\Her\iuRvv\cA\in\C^{J\times J},
\end{align}
where $\iuRvv\coloneqq(\uRnn)^{-1}$ and $\Tk\in\C^{J\times J}$ is thus a full-rank matrix. The centralized MWF can be partitioned as $\mathbf{W}_k =: [\mathbf{W}_{k1}^\T\:\dots\:\mathbf{W}_{kK}^\T]^\T$ where $\mathbf{W}_{kq}$ is applied to $\cy_q$.
Since $\iuRvv$ is block-diagonal with blocks $\{\iuRvv_1,\dots,\iuRvv_K\}$, it holds that:

\begin{equation}\label{eq:Wkq_colSpace}
    \mathbf{W}_{kq} = \iuRvv_q\cA_q\Tk,\:\forall\:(k,q)\in\K^2,
\end{equation}
with $\cA =: [\cA_1^\T\:\dots\:\cA_K^\T]^\T$ similarly to~\eqref{eq:LocalSteeringMat}. 
Now consider the local MWF at node $q$ aiming at estimating $\ld[][q]$ from $\cy_q$, i.e., $\widehat{\mathbf{W}}_q \coloneqq 
    (
        \mathbf{R}_{\mathbf{y}_q\mathbf{y}_q}
    )^{-1}
    \mathbf{R}_{\mathbf{y}_q\mathbf{d}_q}$
where $\mathbf{R}_{\mathbf{y}_q\mathbf{d}_q} \coloneqq \E[{\cy_q\mathbf{d}_q^\Her}]$.
As for~\eqref{eq:mad}, it can be shown that (details omitted):

\begin{equation}\label{eq:hatw}
    \widehat{\mathbf{W}}_q =
    \iuRvv_q
    \cA_q\Tkk[q],
\end{equation}
with $\Tkk[q]\in\C^{J\times J}$ a full-rank matrix.
From~\eqref{eq:Wkq_colSpace} and~\eqref{eq:hatw}, it can be seen that the local MWF at any node $q\in\K$ is in the same column space as the corresponding part of $\mathbf{W}_k$.
The local estimate $\oz[][q] = \widehat{\mathbf{W}}_q^\Her\cy_q$ can then be broadcasted to all other nodes, such that node $k\neq q$ has access to:
\begin{equation}\label{eq:yTilde}
    \oy \coloneqq \begin{bmatrix}
        \ly[\T] & \oz[\T][-k]
    \end{bmatrix}^\T,
\end{equation}
where $\oz[][-k] \coloneqq [\oz[\T][1]\:\dots\:\oz[\T][k-1]\:\oz[\T][k+1]\:\dots\:\oz[\T][K]]^\T$. Node $k$ can compute an MWF to estimate $\ld$ from $\oy$ as:

\begin{equation}\label{eq:wTilde_example_iterationless}
    \oW \coloneqq (\oRyy)^{-1} \oRss \oE \in \C^{\tM \times J},
\end{equation}
where $\oy=:\os+\on$, $\oRyy:=\E[{\oy\oy[\Her]}]$, $\oRyy:=\E[{\os\os[\Her]}]$, and $\oE$ is a selection matrix.
We can partition $\oW$ as $[\mathbf{W}_{kk}^\T\:|\:\mathbf{G}_{k,-k}^\T]^\T$ where $\Wkk\in\C^{M_k\times J}$ and $\Gkmk\in\C^{J(K-1)\times J}$ contains $\Gkq\in\C^{J\times J}$ matrices, $\forall\:q\in\K\backslash\{k\}$. The network-wide parametrization of~\eqref{eq:wTilde_example_iterationless} is then:

\vspace*{-1em}

\begin{equation}\label{eq:nwFilts_example_iterationless}
    \mathbf{W}_k^\mathrm{NW} = [
        (\widehat{\mathbf{W}}_1\mathbf{G}_{k1})^\T\:
        \dots\:
        \mathbf{W}_{kk}^\T\:
        \dots\:
        (\widehat{\mathbf{W}}_K\mathbf{G}_{kK})^\T
    ]^\T,
\end{equation}
where it can be verified that $\oW[\Her]\oy = \mathbf{W}_k^{\mathrm{NW},\Her}\cy$.
Comparing~\eqref{eq:Wkq_colSpace} and~\eqref{eq:hatw}, it can be seen that the centralized filters $\mathbf{W}_k$ are included in the solution space of~\eqref{eq:nwFilts_example_iterationless} and are obtained with $\mathbf{G}_{kq} = \Tkk[q]^{-1}\Tk,\:\forall\:(k,q)\in\K^2$.
No iterations are needed to reach optimality since $\widehat{\mathbf{W}}_k$ is independent of $\oW$.

\subsection{Algorithm definition}\label{subsec:algoDef}

We now consider the complete signal model of~\eqref{eq:centrSigModel}, where $\cN$ and $\uS$ may be non-zero. The iDANSE algorithm can be used to perform optimal one-shot distributed signal estimation in that case. In iDANSE, each node $k$ transmits a $J$-dimensional fused signal $\oz$ which is an estimate of the global component in their local sensor signals.
Each node then uses all the local estimates it receives from other nodes along with its own local sensor signals to reach the centralized solution.

\begin{algorithmic}[1]
    \STATE Define $\gk \coloneqq \Ekk^\T\ccy_{k} \coloneqq \Ekk^\T(\ccs_k + \ccn_k)\in\C^J$.
    \STATE Compute the solution to:
    \vspace*{-.5em}
    \begin{equation}\label{eq:mmseLocal}
        \lcW \coloneqq \underset{\mathbf{W} \in \C^{M_k \times J}}{\mathrm{arg\,min}}\:
        \E[{\|\gk - \mathbf{W}^\Her\cy_k\|^2_2}].
    \end{equation}
    \vspace*{-.5em}
    \STATE Send $\oz \coloneqq \lcW[\Her]\cy_k$ to all other nodes.
    \STATE Receive $\{\oz[][q]\}_{q\in\K\backslash\{k\}}$ and build $\oy$ as~\eqref{eq:yTilde}.
    \STATE Compute $\oW$ via~\eqref{eq:wTilde_example_iterationless}.
    \STATE Obtain desired signal estimate as $\hd \coloneqq \oW[\Her]\oy$.
\end{algorithmic}

The solution to~\eqref{eq:mmseLocal} is the MWF $\lcW = (\lRyy)^{-1} \mathbf{R}_{\mathbf{y}_k{\mathbf{g}}_k}$ where $\lRyy \coloneqq \E[{\cy_k\cy[\Her]_k}]$ and $\mathbf{R}_{\mathbf{y}_k{\mathbf{g}}_k} \coloneqq \E[{\cy_k\gk^\Her}]$.
Since all sources are mutually uncorrelated, $\mathbf{R}_{\mathbf{y}_k{\mathbf{g}}_k} \coloneqq \myglobalaccent{\mathbf{R}}_{\cy_k\cy_k}\Ekk$ with $\myglobalaccent{\mathbf{R}}_{\cy_k\cy_k} \coloneqq \E[{\ccy_k\ccy[\Her]_k}]$.
The SCMs $\lRyy$ and $\myglobalaccent{\mathbf{R}}_{\cy_k\cy_k}$ must be estimated from the data, which can be achieved using a VAD distinguishing between the contributions of local and global sources~\cite{zhao2020model, dov2017multimodal, kim2018voice}. Note that this VAD estimation problem may be less straightforward to solve than the traditional case encountered in DANSE~\cite{bertrand2010distributed}.

\subsection{iDANSE optimality and comparison with DANSE}\label{subsec:optimalityCriterion}

In iDANSE, the fused signals $\oz$ must be of sufficiently high dimension to span the latent subspace formed by the global source signals $\ccs[\mathrm{lat}]$ and $\ccn[\mathrm{lat}]$, i.e., $J\geq \cS+\cN$.
In that case, iDANSE is optimal in the general case~\eqref{eq:centrSigModel} as formulated in Theorem 1 (proof is an extension of~\secref{subsec:oneshotexample}, omitted due to space constraints). Note that this differs from~\cite{plata2015distributed}, which states that $J$ must be at least equal to the dimension of the latent desired subspace spanned by the entire WASN.

\vspace*{-1.25em}

\noindent
\paragraph*{Theorem 1.} \textit{
    Let $\ld=\cA_k\ccs[\mathrm{lat}] + \uA_k\cus[\mathrm{lat}]$ for all $k\in\K$ with $\ccs[\mathrm{lat}]$ and $\cus[\mathrm{lat}]$ complex $\cS$-channel and $\uS$-channel signals, respectively, $\cA_k$ a $J\times \cS$ matrix of rank $\cS$, and $\uA_k$ a $J\times \uS$ matrix of rank $\uSk<\uS$. Let there be $\cN$ global noise sources in the environment. If $J \geq \cS + \cN$, then the iDANSE algorithm reaches the centralized LMMSE solution~\eqref{eq:mwfCentr} for any $k$.
}

\vspace*{1em}

If $\uS{=}0$, both DANSE and iDANSE reach the centralized solution. A trade-off then appears between convergence speed and bandwidth usage, as DANSE requires $\cS$ fused signals but multiple iterations to converge, while iDANSE reaches the optimum in a single cycle but requires $\cS {+} \cN$ fused signals.

\vspace*{-.85em}

\section{Numerical results}
\label{sec:res}

\subsection{Convergence assessment}\label{subsec:res_convergence}

\begin{figure}[h]
    \centering
    \includegraphics[width=\columnwidth,trim={0 1.5em 0 0},clip=false]{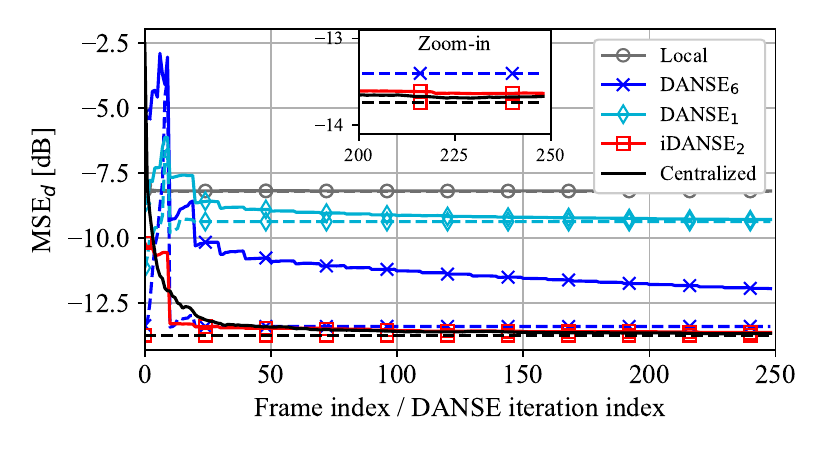}
    \caption{MSE$_d$ as function of time frame index (iteration index for DANSE). True SCMs: dashed lines. Online SCMs estimates: solid lines.}
    \label{fig:res}
\end{figure}

In order to assess the convergence of iDANSE, a general case where latent signal samples and steering matrix entries are drawn from standard normal distributions is first considered.
A fully connected WASN with $K{=}10$ nodes and $M_k{=}8,\:\forall\:k$ is deployed in a static scene composed of $\cS{=}1$ global desired source, $\uSk{=}5,\:\forall\:k$ local desired sources, $\cN{=}1$ global noise source, and $\uNk{=}3,\:\forall\:k$ local noise sources. All localized sources are given the same power to obtain a signal-to-noise ratio (SNR) of 0 dB on average over all nodes.
Sensor noise is scaled to 10\% of the first sensor signal power at all nodes.
The performance of (i) iDANSE is compared to that of (ii) centralized processing, (iii) DANSE with sequential node-updating using $J=\cS+\mathrm{max}_k\uSk$ or (iv) $J=\cS$ and (v) processing using only the $M_k$ local sensor signals. The mean square error (MSE) defined as $\mathrm{MSE}_d[l] = \frac{1}{KJB}\sum_{k\in\K}\sum_{n=lB}^{(l+1)B-1}
\|
    \hat{\mathbf{d}}_k[n] - \mathbf{d}_k[n]
\|_{2}^2$ is computed between the true desired signals and the estimates obtained via each of the mentioned algorithms. This is first done once using the true SCMs computed from the steering matrices themselves. To simulate online-mode processing, frame-based SCM estimation via exponential averaging is employed by frame size $B=250$ samples and forgetting factor 0.995. The value of $J$ is indicated as $x$ in legend entries of the type (i)DANSE$_x$. The results in~\figref{fig:res} show that iDANSE reaches the centralized performance instantly with true SCMs and very quickly in online-mode while DANSE is unable to reach the optimum, even with $J=\cS+\mathrm{max}_k\uSk$.

\vspace*{-.5em}

\subsection{Speech enhancement scenario}

\begin{figure}[h]
    \centering
    \begin{scriptsize}
        \input{metrics_fixed_2.tex}
    \end{scriptsize}
    \vspace*{-1em}
    \caption{eSTOI and SNR averages over all nodes.}
    \label{fig:res_snr}
\end{figure}
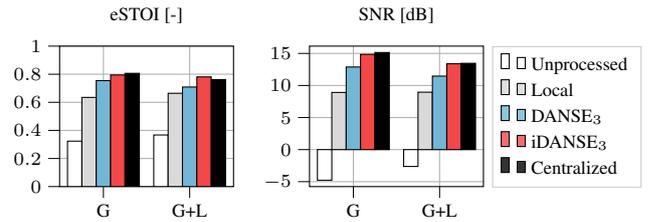

Consider a scenario where a fully connected WASN with $K{=}3$ nodes $M_k{=}6,\:\forall\:k$ is deployed in a static scene with a reverberation time of 0.15\,s. The node-specific target signals include a speech signal produced by a single PA system ($\cS{=}1$) and may (scenario ``G'': $\uSk{=}1,\:\forall\:k$) or may not (scenario ``G+L'': $\uSk{=}0,\:\forall\:k$) include a local desired speech source. Speech signals consist of LibriSpeech~\cite{panayotov2015librispeech} snippets. The scene includes $\cN{=}2$ global and $\uNk{=}1,\:\forall\:k$ local babble noise sources. Processing is done over 10\,s of signal using the weighted overlap-add (WOLA) framework~\cite{didier2023sampling, bertrand_adaptive_2009} with 1024 samples per frame, 50\% frame-overlap, and a forgetting factor of 0.995. The performance is evaluated in terms of extended short-term objective intelligibility (eSTOI)~\cite{jensen2016algorithm} and SNR of the first desired signal estimate channel on average over all nodes, for algorithms (i), (ii), (iii), and (v) -- as defined in~\secref{subsec:res_convergence}.
The results in \figref{fig:res_snr} show that iDANSE outperforms DANSE in both ``G'' and ``G+L'' scenarios. Indeed, even when $\uSk{=}0,\:\forall\:k$, a better performance is achieved by iDANSE due to its faster convergence.

\vspace*{-1em}

\vspace*{-2em}

\section{Conclusions}
\label{sec:ccl}

In this paper, the iDANSE algorithm has been introduced for iterationless LMMSE estimation of node-specific signals in a fully connected WASN where latent desired subspaces may be non-overlapping. Future work will address the generalization of iDANSE to partially overlapping latent subspaces and a topology-independent iDANSE formulation.

\bibliographystyle{IEEEbib}
\bibliography{mybib}

\end{document}

%% file: metrics_fixed_2.tex
\begin{tikzpicture}

\definecolor{darkgray176}{RGB}{176,176,176}
\definecolor{gainsboro217}{RGB}{217,217,217}
\definecolor{lightgray204}{RGB}{204,204,204}
\definecolor{skyblue114183214}{RGB}{114,183,214}
\definecolor{tomato2447272}{RGB}{244,72,72}

\begin{groupplot}[group style={group size=2 by 1}]
  
\nextgroupplot[
  tick align=outside,
  tick pos=left,
  height=14em,
  width=.45\columnwidth,
  x grid style={darkgray176},
  xmajorgrids,
  xmin=-0.508333333333333, xmax=1.50833333333333,
  xtick style={color=black},
  xtick={0,1},
  xticklabels={G,G+L},
  y grid style={darkgray176},
  title={eSTOI [-]},
  ymajorgrids,
  ymin=0, ymax=1,
  ytick style={color=black}
  ]
  \draw[draw=black,fill=white] (axis cs:-0.416666666666667,0) rectangle (axis cs:-0.25,0.323518845861445);
  \draw[draw=black,fill=gainsboro217] (axis cs:-0.25,0) rectangle (axis cs:-0.0833333333333333,0.634867753010975);
  \draw[draw=black,fill=skyblue114183214] (axis cs:-0.0833333333333333,0) rectangle (axis cs:0.0833333333333333,0.754518469662928);
  \draw[draw=black,fill=tomato2447272] (axis cs:0.0833333333333333,0) rectangle (axis cs:0.25,0.794528217808634);
  \draw[draw=black,fill=black] (axis cs:0.25,0) rectangle (axis cs:0.416666666666667,0.806286663198023);
  \draw[draw=black,fill=white] (axis cs:0.583333333333333,0) rectangle (axis cs:0.75,0.367343229168073);
  \draw[draw=black,fill=gainsboro217] (axis cs:0.75,0) rectangle (axis cs:0.916666666666667,0.664321202429806);
  \draw[draw=black,fill=skyblue114183214] (axis cs:0.916666666666667,0) rectangle (axis cs:1.08333333333333,0.708745469578965);
  \draw[draw=black,fill=tomato2447272] (axis cs:1.08333333333333,0) rectangle (axis cs:1.25,0.780959541244599);
  \draw[draw=black,fill=black] (axis cs:1.25,0) rectangle (axis cs:1.41666666666667,0.761571751741221);
  
\nextgroupplot[
  legend cell align={left},
  legend style={
    fill opacity=0.8,
    draw opacity=1,
    text opacity=1,
    at={(1.05,1)},
    anchor=north west,
    draw=lightgray204
  },
tick align=outside,
tick pos=left,
height=14em,
width=.45\columnwidth,
x grid style={darkgray176},
xmajorgrids,
xmin=-0.508333333333333, xmax=1.50833333333333,
xtick style={color=black},
xtick={0,1},
xticklabels={G,G+L},
y grid style={darkgray176},
title={SNR [dB]},
ymajorgrids,
ymin=-5.76797657626503, ymax=16.1372385328193,
ytick style={color=black}
]
\draw[draw=black,fill=white] (axis cs:-0.416666666666667,0) rectangle (axis cs:-0.25,-4.77228498039756);
\addlegendimage{ybar,ybar legend,draw=black,fill=white}
\addlegendentry{Unprocessed}

\draw[draw=black,fill=gainsboro217] (axis cs:-0.25,0) rectangle (axis cs:-0.0833333333333333,8.91316260815648);
\addlegendimage{ybar,ybar legend,draw=black,fill=gainsboro217}
\addlegendentry{Local}

\draw[draw=black,fill=skyblue114183214] (axis cs:-0.0833333333333333,0) rectangle (axis cs:0.0833333333333333,12.8817173724766);
\addlegendimage{ybar,ybar legend,draw=black,fill=skyblue114183214}
\addlegendentry{DANSE$_3$}

\draw[draw=black,fill=tomato2447272] (axis cs:0.0833333333333333,0) rectangle (axis cs:0.25,14.8456816013894);
\addlegendimage{ybar,ybar legend,draw=black,fill=tomato2447272}
\addlegendentry{iDANSE$_3$}

\draw[draw=black,fill=black] (axis cs:0.25,0) rectangle (axis cs:0.416666666666667,15.1415469369518);
\addlegendimage{ybar,ybar legend,draw=black,fill=black}
\addlegendentry{Centralized}

\draw[draw=black,fill=white] (axis cs:0.583333333333333,0) rectangle (axis cs:0.75,-2.62838921584039);
\draw[draw=black,fill=gainsboro217] (axis cs:0.75,0) rectangle (axis cs:0.916666666666667,8.9532079143407);
\draw[draw=black,fill=skyblue114183214] (axis cs:0.916666666666667,0) rectangle (axis cs:1.08333333333333,11.4630861182975);
\draw[draw=black,fill=tomato2447272] (axis cs:1.08333333333333,0) rectangle (axis cs:1.25,13.3840213031707);
\draw[draw=black,fill=black] (axis cs:1.25,0) rectangle (axis cs:1.41666666666667,13.4572863561828);
\end{groupplot}

\end{tikzpicture}